# Ion-Beam Modification of Metastable Gallium Oxide Polymorphs

D. I. Tetelbaum[1,a], A. A. Nikolskaya[1], D. S. Korolev[1], A. I. Belov[1], V. N. Trushin[1], Yu. A. Dudin[1], A. N. Mikhaylov[1], A. I. Pechnikov[2,3], M. P. Scheglov[2], V. I. Nikolaev[2,3], D. Gogova[1,4]

[1]*Research Institute of Physics and Technology, Lobachevsky University, Nizhny Novgorod 603022, Russia*
[2]*Ioffe Institute, Russian Academy of Sciences, St. Petersburg, 194021 Russia*
[3]*Perfect Crystals LLC, Saint-Petersburg, 194064 Russia*
[4]*University of Oslo, Blindern, NO-0316 Oslo, Norway*

**Abstract.** Gallium oxide with a corundum structure ($\alpha$-$Ga_2O_3$) has recently attracted great attention in view of electronic and photonic applications due to its unique properties including a wide band gap exceeding that of the most stable beta phase ($\beta$-$Ga_2O_3$). However, the lower thermal stability of the $\alpha$-phase at ambient conditions in comparison with the $\beta$-phase requires careful investigation of its resistance to other external influences such as ion irradiation, ion doping, etc. In this work, the structural changes under the action of $Al^+$ ion irradiation have been investigated for a polymorphic gallium oxide layers grown by hydride vapor phase epitaxy on c-plane sapphire and consisting predominantly of $\alpha$-phase with inclusions of $\varepsilon(\kappa)$-phase. It is established by the X-ray diffraction technique that inclusions of $\varepsilon(\kappa)$-phase in the irradiated layer undergo the expansion along the normal to the substrate surface, while there is no a noticeable deformation for the $\alpha$-phase. This speaks in favor of the different radiation tolerance of various $Ga_2O_3$ polymorphs, especially the higher radiation tolerance of the $\alpha$-phase. This fact should be taken into account when utilizing ion implantation to modify gallium oxide properties in terms of development of efficient doping strategies.

---

[a] Electronic mail: tetelbaum@phys.unn.ru

Recently, there has been a continuously growing interest in gallium oxide polymorphs due to the need of wide band gap semiconductors for the next generation power electronic devices, ultraviolet detectors (including solar-blind detectors) and semiconductor devices capable of operating in harsh environments and space [1-5]. There are several crystalline modifications of this semiconductor, among which the β-phase is the most stable and studied in detail. On the basis of this phase, fabrication technologies of a number of semiconductor devices have been developed and demonstrated. However, more and more scientific attention has been attracted recently by the other polymorphs of $Ga_2O_3$, in particular by the α-$Ga_2O_3$ phase. This is because the α-phase has a wider band gap ($E_g$ = 5.2 eV) compared to the β-phase (4.5 – 4.9 eV) and therefore, it possesses potentially a higher breakdown electrical field value. In addition, the α-phase of $Ga_2O_3$ has a hexagonal crystalline structure (in contrast to the low-symmetry monoclinic one of the β-phase) like the readily available and easy to handle sapphire substrate used for cost-effective heteroepitaxial growth of gallium oxide layers. Therefore, the α-phase of $Ga_2O_3$ has a better lattice parameter matching to the α-$Al_2O_3$, which makes it possible to improve the epilayers structural quality.

Ion implantation is a traditional powerful technology for device processing in semiconductor electronics, and has already been extensively employed in the case of the thermodynamically stable β-$Ga_2O_3$ phase (see for example Ref. [6-9]). However, to the best of our knowledge, there are no similar studies for the α-$Ga_2O_3$.

In this report, the effect of $Al^+$ ion irradiation on the structure of two polymorphic (α + ε(κ)) $Ga_2O_3$ layers grown by hydride vapor phase epitaxy (HVPE) has been demonstrated for the first time.

The polymorphic $Ga_2O_3$ layers were epitaxially grown on α-$Al_2O_3$ (001) substrates by the industry-relevant HVPE method at LLC Perfect Crystals. A hot-wall atmospheric pressure reactor was employed providing homogenous temperature distribution and high growth rates. In the deposition process gaseous HCl flows over metallic gallium forming gallium chloride vapor, which reacts with oxygen on the substrate heated to ~ 500 °C (below the transition point from α- to β-phase gallium oxide). Argon was employed as a carrier gas. The thickness of the epilayers was determined as ~ 1 μm. Irradiation with $Al^+$ was carried out on a Raduga-3M implanter [10-12]. In this implanter, pulsed ion beams were generated by sputtering of Al metal foil with a plasma of electrical discharge that occurred between the foil and the grounded electrode at a field strength exceeding a certain threshold value. After sputtering, the $Al^+$ ions, emitted from the foil surface, were accelerated by an electric field up to an energy of 30 keV. The fluence was approximately $3 \cdot 10^{15}$ $cm^{-2}$. According to the SRIM calculations [13], the mean projected range ($R_p$) was 25 nm, and the straggling ($\Delta R_p$) – 12 nm.

The structure and composition of the as-grown and irradiated epilayer were investigated by X-ray diffraction measurements employing a Shimadzu XRF-7000 diffractometer equipped with an X-ray tube with a copper anode. The ω – 2θ beam scanning was performed in the 2θ range from 15 to 70 degrees.

Fig. 1 illustrates diffraction patterns typical of the HVPE gallium oxide epilayers before and after $Al^+$ ion irradiation. Notably, before irradiation, the diffraction pattern is composed of Bragg reflections from the α-phase $Ga_2O_3$ with indices (006) in epitaxial relationship with the (006) peak of the c-plane sapphire substrate. In addition, some low intense diffraction lines appear, which can be identified as reflections from the ε- or κ-phase of $Ga_2O_3$ [14-16]. The instrumental resolution does not allow the separation of the very close lines of these two phases, therefore, following the conclusions of Ref. [15], we assign these lines to Bragg reflexes from the ε(κ)-phase. (The question about the separate identification and interpretation of the ε (hexagonal) and κ (orthorhombic) phases is still under debate [16]. Since the XRD peaks related to the ε(κ)-phase are significantly weaker in intensity than these of the α-phase, obviously, the volume fraction of the ε(κ)-phase is rather small in our gallium oxide heteroepitaxial layers.

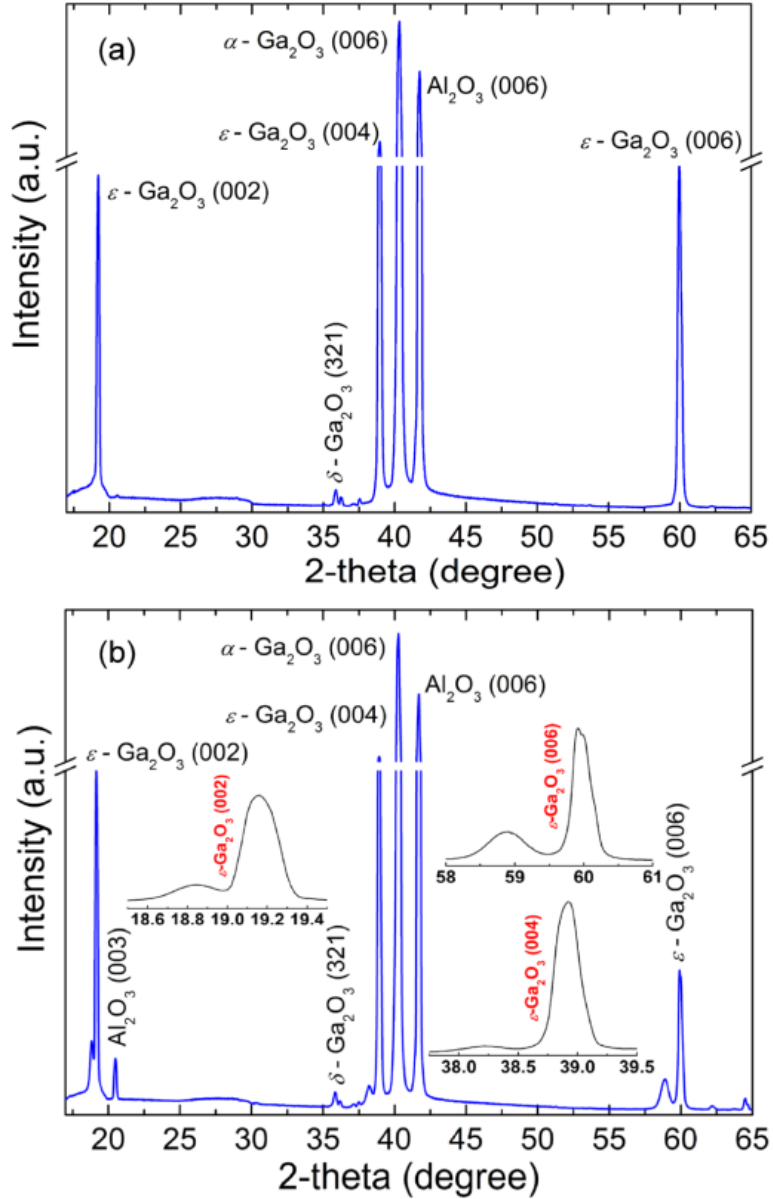

FIG. 1. Typical X-ray ω – 2θ scans of the as-grown HVPE Ga$_2$O$_3$/sapphire epilayer (a) and of the epilayer irradiated with Al$^+$ ions (b). The insets in Figure 1b depicts the enlarged XRD diffraction lines of the ε(κ)-phase.

Evidently, after ion irradiation, the XRD peaks of the α-phase and ε(κ)-phase are preserved as intensity and peak positions, however, some new lines appear near the lines of the ε(κ)-phase. The question is, whether the appearance of these lines is associated with a phase transformation (see e.g. [17]) or with the deformation of the ε(κ)-phase inclusions. To answer this question, the relative change in the interplanar distance $\Delta d/d$ has been calculated for the newly emerged lines relative to the nearest preserved lines of the ε(κ)-phase with indices (002), (004) and (006), using the relation $2d\sin(\theta) = n\lambda$. It is found out that the $\Delta d/d$ values for these lines are nearly the same ~ 0.02. This fact is *indeed* consistent with the assumption that the observed additional lines in the diffraction patterns are due to the deformation of the ε(κ)-phase. The sign of deformation corresponds to the lattice expansion along the normal to the sample surface. This expansion might be associated with the incorporation of interstitial atoms in the lattice.

An alternative assumption is that the irradiation has induced the ε(κ) → β phase transition, in analogy with the β → κ transition observed by Anber *et al.* in Ref. [17]. However, a detailed analysis of the new lines positions shows that the deviation from the tabulated values is rather large making this mechanism less probable.

It is interesting that, in contrast to the ε(κ)-phase, the diffraction line of the initial α-phase does not undergo a significant shift after irradiation. This indicates a higher radiation tolerance of the α-phase in comparison with the ε(κ)-phase. It should be also noted that no line shift was observed also in Ref. [18] for the β-$Ga_2O_3$ phase upon irradiation with high-energy oxygen ions.

In conclusion, the ion-induced modification of $Ga_2O_3$ layers with various polymorphs was revealed. A detailed analysis of radiation impact on different $Ga_2O_3$ polytypes as well as a deep understanding of the origin of their different radiation tolerance requires further investigations employing additional analytical techniques and is underway.

**ACKNOWLEDGMENTS**
This study was supported by the Lobachevsky University competitiveness program in the frame of 5-100 Russian Academic Excellence Project.

**DATA AVAILABILITY**
The data that support the findings of this study are available from the corresponding author upon reasonable request.

**REFERENCES**
[1] E. Ahmadi and Y. Oshima, J. Appl. Phys. **126**, 160901 (2019).
[2] S.J. Pearton, J. Yang, P.H. Cary, F. Ren, J. Kim, M.J. Tadjer, and M.A. Mastro, Appl. Phys. Rev. **5**, 011301 (2018).
[3] S.J. Pearton, F. Ren, M. Tadjer, and J. Kim, J. Appl. Phys. **124**, 220901 (2018).
[4] M. Higashiwaki and G.H. Jessen, Appl. Phys. Lett. **112**, 060401 (2018).
[5] M.A. Mastro, A. Kuramata, J. Calkins, J. Kim, F. Ren, and S.J. Pearton, ECS J. Solid State Sci. Technol. **6**, 356 (2017).
[6] K. Sasaki, M. Higashiwaki, A. Kuramata, T. Masui, and S. Yamakoshi, Appl. Phys. Express **6(8)**, 086502 (2013).
[7] E. Wendler, E. Treiber, J. Baldauf, S. Wolf, and C. Ronning, Nucl. Instruments Methods Phys. Res. Sect. B Beam Interact. with Mater. Atoms **379**, 85 (2016).
[8] M.H. Wong, H. Murakami, Y. Kumagai, and M. Higashiwaki, IEEE Electron Device Lett. **41**, 296 (2020).
[9] A. Nikolskaya, E. Okulich, D. Korolev, A. Stepanov, D. Nikolitchev, A. Mikhaylov, D. Tetelbaum, A. Almaev, C.A. Bolzan, A.J. Buaczik, R. Giulian, P. Luis, A. Kumar, M. Kumar, D. Gogova, J. Vacuum Sci. Technol. A. **39**, (2021) In press
[10] A.I. Ryabchikov, N.M. Arzubov, S.V. Dektyarev, Instrum Exp Tech. **34**, 173 (1991)
[11] A.I. Ryabchikov, I.B. Stepanov, S.V. Dektyarev, E.I. Lukonin, A.I. Shulepov, Rev. Sci. Instrum. **71(2)**, 704 (2000)
[12] A.I. Ryabchikov, I.A. Ryabchikov, I.B. Stepanov, S.V. Dektyarev. Rev. Sci. Instrum. **77(3)**, 03B516 (2006)

[13] J.F. Ziegler, M.D. Ziegler, J.P. Biersack, Nuclear Instrum. Methods Phys. Research B: Beam Interactions with Materials and Atoms, **268**, 1818 (2010)

[14] R. Roy, V.G. Hill, E.F. Osborn, J. Am. Chem. Soc. **74**, 719 (1952)

[15] V.D. Wheeler, N. Nepal, D.R. Boris, S.B. Qadri, L.O. Nyakiti, A. Lang, A. Koehler, G. Foster, S.G. Walton, C.R. Eddy, D.J. Meyer, Chem. Mater. **32**, 1140 (2020)

[16] I. Cora, F. Mezzadri, F. Boschi, M. Bosi, M. Čaplovičová, G. Calestani, I. Dódony, B. Pécz, R. Fornari, Cryst. Eng. Comm. **19**, 1509 (2017)

[17] E.A. Anber, D. Foley, A.C. Lang, J. Nathaniel, J.L. Hart, M.J. Tadjer, K.D. Hobart, S. Pearton, M.L. Taheri, Appl. Phys. Lett. **117**, 152101 (2020)

[18] C. Liu, Y. Berencén, J. Yang, Y. Wei, M. Wang, Y. Yuan, C. Xu, Y. Xie, X. Li, S. Zhou, Semicond. Sci. Technol. **33**, 095022 (2018)